\documentclass[showpacs,twocolumn,prl,aps,color]{revtex4}
\usepackage{graphicx}
\begin{document}

\title{Variational study of the quantum phase transition in bilayer Heisenberg model with Bosonic RVB wave function}

\author{Haijun Liao and Tao Li}
\affiliation{Department of Physics, Renmin University of China,
Beijing, 100872, P.R.China}
\date{\today}

\begin{abstract}
We study the ground state phase diagram of the bilayer Heisenberg
model on square lattice with a Bosonic RVB wave function. The wave
function has the form of a Gutzwiller projected Schwinger Boson mean
field ground state and involves two variational parameters. We find
the wave function provides an accurate description of the system on
both sides of the quantum phase transition. Especially, through the
analysis of the spin structure factor, ground state fidelity
susceptibility and the Binder moment ratio $Q_{2}$, a continuous
transition from the antiferromagnetic ordered state to the quantum
disordered state is found at the critical coupling of
$\alpha_{c}=J_{\perp}/J_{\parallel}\approx2.62$, in good agreement
with the result of quantum Monte Carlo simulation. The critical
exponent estimated from the finite size scaling
analysis($1/\nu\approx1.4$) is consistent with that of the classical
3D Heisenberg universality class.
\end{abstract}

\maketitle
\section{I. Introduction}
The study of quantum phase transition is a central issue in modern
condensed matter physics. It is widely believed that the
Ginzburg-Landau-Wilson theory for classical phase transition may
fail to describe the quantum phase transition as a result of the
quantum interference effect between classical paths(the Berry phase
effect). Recently, the concepts of quantum order and de-confined
quantum criticality are put forward theoretically. On the
experimental side, the study of quantum phase transition plays an
important role in areas ranging from high-Tc cuprates, heavy Fermion
systems, to the cold atom systems\cite{Sachdev1}.

The bilayer Heisenberg model(BHM) on square lattice is a standard
model for the study of quantum phase transition. With the increase
of the interlayer coupling($J_{\perp}$) over the intralayer
coupling($J_{\parallel}$), the ground state of the system evolves
from a state with antiferromagnetic long rang order to a quantum
disordered state through a continuous phase transition. Much
theoretical and numerical efforts have been devoted to the study of
this quantum phase transition.

On theoretical side, perturbative calculations starting from both
the ordered side(spin wave expansion)\cite{spinwave} and the
disordered side(the bond operator expansion)\cite{bond} have been
applied to the system. However, due to the biased nature of
pertubative methods, none of them can give an accurate description
of the system in the near vicinity of the quantum phase transition.
The problem is also treated with the Schwinger Boson mean field
theory\cite{schwinger,Millis}. Although the theory does predict a
phase transition between the antiferromagnetic ordered state and the
quantum disordered state, the nature of the transition is incorrect.
The mean field theory predicts a discontinues dimerization
transition around $J_{\perp}/J_{\parallel}=4.62$ into a state
composed of independent interlayer dimers, while in the real system,
the intralayer correlation is nonzero for any finite
$J_{\perp}/J_{\parallel}$.

On numerical side, the model is thoroughly studied by a variety of
methods including the high temperature series expansion\cite{series}
and the quantum Monte Carlo simulation(Stochastic series
expansion)\cite{qmc,Wang}. These numerical works confirm the
existence of the quantum critical point around
$\alpha=J_{\perp}/J\approx 2.52$. The critical exponents is found to
be consistent with that of the classical 3D Heisenberg universality
class, indicating the irrelevance of the Berry phase effect in this
phase transition.

As the quantum phase transition occurs at zero temperature, it is
natural to find a description of it in terms of an explicit ground
state wave function. The variational approach to quantum phase
transition has the virtue that it focus directly on the zero
temperature behavior of the system and provides much more detailed
information on the quantum critical behavior. In this regard, a
RVB-type variational wave function\cite{RVB,loop} had been applied
to the study of the quantum phase transition in the BHM. The wave
function is derived from Gutzwiller projection of Schwinger Boson
mean field ground state. It is well known that such a RVB wave
function can describe both the magnetic ordered and the quantum
disordered state. Thus, it has the potential to provide an unbiased
description of the quantum phase transition in BHM. The same type of
variational wave function has been successfully applied to the study
of the single layer two-dimensional Heisenberg
model\cite{RVB,schwinger1}. However, for the BHM, the variational
calculation in \cite{Yoshioka} using such a wave function predicts a
critical coupling $\alpha_{c}=3.51$, which is a very bad estimate as
compared to the result of numerical simulation. A central issue to
be addressed in this paper is to understand why the Bosonic RVB
state, which works so well on square lattice, fails for the BHM and
how to improve it.

In this paper, we propose a RVB-type variational wave function with
two variational parameters for the BHM. Similar to \cite{Yoshioka},
our wave function is derived from Gutzwiller projected Schwinger
Boson mean field state. However, in our theory the intralayer RVB
pairing and interlayer RVB pairing are treated as two independent
variational parameters, rather than been determined by mean field
self-consistent equations. We find our variational wave function
provides an accurate description of the quantum phase transition of
the BHM. We find the the transition is continuous. By analyzing the
spin structure factor, ground state fidelity susceptibility and
Binder moment ratio $Q_{2}$, the critical coupling strength is
estimated to be $\alpha_{c}\approx 2.62$, in good agreement with
those determined from the numerical simulation. The critical
exponent estimated from the scaling analysis of the $Q_{2}$ data is
also consistent wit that of the classical 3D Heisenberg universality
class. Our result indicates that the Bosonic RVB wave function
derived from Gutzwiller projection of the Schwinger Boson mean field
state can provides accurate description of the quantum phase
transition in quantum antiferromagnets. We also find that the
failure of the Schwinger Boson mean field theory originates from the
overestimation of the tendency to form interlayer dimers, which is
again caused by the relaxation of the no double occupancy
constraint.

The paper is organized as follows. In section II, we introduce the
BHM and the Bosonic RVB wave function. In section III, we present
the numerical method to do calculation on such wave functions. In
section IV, we present the numerical results and determine the
critical point of the phase transition by analyzing the results of
fidelity susceptibility and Binder moment ratio. In section V, we
present a discussion on related issues and conclude this paper.

\section{II. The bilayer Heisenberg model and the RVB-type variational wave function}

The model(BHM) we study in this paper is given by
\begin{equation}
\mathrm{H}=J_{\parallel}\sum_{<i,j>,\mu}\vec{\mathrm{S}}_{i}^{\mu}\cdot\vec{\mathrm{S}}_{j}^{\mu}
+J_{\perp}\sum_{i}\vec{\mathrm{S}}_{i}^{1}\cdot\vec{\mathrm{S}}_{i}^{2},
\end{equation}
where $\vec{\mathrm{S}}_{i}^{\mu}$ denotes the spin operator at site
$i$ of layer $\mu(=1,2)$. $\sum_{<i,j>}$ means the summation over
nearest-neighboring sites on the square lattice of each layer.
$\alpha=J_{\perp}/J_{\parallel}$ is the only dimensionless parameter
of the model. When $\alpha=0$, the model describes two decoupled
two-dimensional Heisenberg model, each of which are
antiferromagnetic ordered at zero temperature. When $\alpha
\rightarrow \infty $, the system reduces to N decoupled interlayer
dimers and the system is in a trivial quantum disordered state. A
continuous quantum phase transition connects these two limits.
Earlier numerical simulation shows that the phase transition occurs
around $\alpha_{c}=2.52$\cite{qmc,Wang}.

The Bosonic RVB wave function we will adopt in this study is made of
coherent superposition of spin singlet configurations on the lattice
and can be written as
\begin{equation}
    |\mathrm{RVB}\rangle=\sum_{\{i_{k},j_{k}
    \}}A(\{i_{k},j_{k}\})\prod_{k=1}^{N/2}S(i_{k},j_{k}),
\end{equation}
in which
$S(i_{k},j_{k})=\frac{1}{\sqrt{2}}(\uparrow_{i_{k}}\downarrow_{j_{k}}-\downarrow_{i_{k}}\uparrow_{j_{k}})$
denotes the spin singlet pair between site $i_{k}$ and $j_{k}$.
$A(\{i_{k},j_{k}\})$ are the coefficients of the coherent
superposition. In our case, $A(\{i_{k},j_{k}\})$ can be written in a
factorizeable
 form $A(\{i_{k},j_{k}\})=\prod_{k=1}^{N/2}a_{i_{k},j_{k}}$

The wave function Eq.(2) can be used directly as variational state
for quantum antiferromagnet. A more efficient and intuitively more
attractive way to generate the RVB wave function is by Gutzwiller
projection of Schwinger Boson mean field state. This approach is
used to study two-dimensional Heisenberg model and is proved to be
very successful. However, direct application of the approach to the
BHM results in unsatisfactory results.

Here, we will adopt the form of the Gutzwiller projected Schwinger
Boson mean field state, but regard the mean field order
parameters(intralayer and interlayer RVB pairing amplitudes) as free
variational parameters, rather than been determined from the mean
field self-consistent equations. The reason for such a choice is as
follows. In the mean field treatment, the no double occupancy
constraint is relaxed. As a result, the quantitative prediction of
the mean field theory is not reliable. For example, the mean field
equation predicts an un-physical dimerization transition for BHM at
$\alpha\approx 4.62$, whose origin can be traced back to the
overestimation the tendency to form interlayer dimers, which is
again related to the  relaxation of the local constraint.

In the Schwinger Boson representation\cite{schwinger}, the spin
operator is written as
\begin{equation}
    \vec{\mathrm{S}}=\frac{1}{2}\sum_{\alpha,\beta=1,2}b^{\dagger}_{\alpha}\vec{\sigma}_{\alpha,\beta}b_{\beta},
\end{equation}
in which $b_{\alpha}$ is a Boson operator, $\vec{\sigma}$ is the
Pauli matrix. Eq.(3) is a faithful representation of the spin
algebra provided that the Bosonic particle satisfy the no double
occupancy constraint
\begin{equation}
    \sum_{\alpha}b^{\dagger}_{\alpha}b_{\alpha}=1.
\end{equation}
The BHM written in terms of the Schwinger Boson operators reads
\begin{eqnarray}
    \mathrm{H}=&-&J_{\parallel}\sum_{<i,j>,\mu}\hat{\Delta}_{i,j}^{\mu
    \dagger}\hat{\Delta}_{i,j}^{\mu}-J_{\perp}\sum_{i}\hat{\Delta}_{i}^{\dagger}\hat{\Delta}_{i}\nonumber \\
    &-&\sum_{i,\mu}\lambda_{i,\mu}(n_{i,\mu}-1)
\end{eqnarray}
in which
\begin{eqnarray}
\hat{\Delta}_{i,j}^{\mu}&=&\frac{1}{\sqrt{2}}(b_{i,\mu,\uparrow}b_{j,\mu,\downarrow}-b_{i,\mu,\downarrow}b_{j,\mu,\uparrow}) \nonumber\\
\hat{\Delta}_{i}&=&\frac{1}{\sqrt{2}}(b_{i,1,\uparrow}b_{i,2,\downarrow}-b_{i,1,\downarrow}b_{i,2,\uparrow})
\end{eqnarray}
denote the intralayer and interlayer RVB pairing operator,
$n_{i,\mu}=\sum_{\alpha=\uparrow,\downarrow}b_{i,\mu,\alpha}^{\dagger}b_{i,\mu,\alpha}$.
The Largrange multiplier $\lambda_{i,\mu}$ is introduced to keep
track of the local constraint.

In the mean field theory, we treat $\lambda_{i,\mu}=\lambda$ as a
constant and decouple the interaction term using the following mean
field order parameters
$\Delta_{\parallel}=\langle\hat{\Delta}_{i,j}^{1}\rangle=\langle\hat{\Delta}_{i,j}^{2}\rangle
$ and $ \Delta_{\perp}=\langle\hat{\Delta}_{i}\rangle$. The mean
field Hamiltonian reads(up to a constant)
\begin{eqnarray}
    \mathrm{H}_{\mathrm{MF}}&=&-J_{\parallel}\Delta_{\parallel}\sum_{<i,j>,\mu}(\hat{\Delta}_{i,j}^{\mu
    \dagger}+\hat{\Delta}_{i,j}^{\mu})\nonumber \\
    &&-J_{\perp}\Delta_{\perp}\sum_{i}(\hat{\Delta}_{i}^{\dagger}+\hat{\Delta}_{i})-\lambda\sum_{i,\mu}n_{i,\mu}.
\end{eqnarray}

The mean field ground state of Eq.(7) reads
\begin{eqnarray}
    |G\rangle =
    \exp\Bigl[\sum_{i,j;\mu,\nu}&a_{i\mu,j\nu}&(b_{i\mu,\uparrow}^{\dagger}b_{j\nu,\downarrow}^{\dagger}-b_{i\mu,\downarrow}^{\dagger}b_{j\nu,\uparrow}^{\dagger})
     \Bigr] \Bigl|0\Bigr\rangle,
\end{eqnarray}
in which $|0\rangle$ denotes the vacuum of the Schwinger Boson.
$a_{i\mu,j\nu}$ represents the RVB amplitude between site $i$ in
$\mu$ layer and site $j$ in $\nu$ layer. As a result of the
bipartite nature of the system, the RVB amplitude is nonzero only
for sites belonging to different sublattices. Thus for $\mu=\nu$,
$a_{i\mu,j\nu}$ is nonzero only when $i,j$ have different parity,
while for $\mu\neq\nu$ the reverse is true. The intralayer and
interlayer RVB amplitudes are given
by($a_{i1,j1}=a_{i2,j2},a_{i1,j2}=a_{i2,j1}$ by symmetry)
\begin{eqnarray}
    a_{i1,j1}=\frac{1}{N}\sum_{\vec{k}}[\xi(k)+\eta(k)]\exp(i\vec{k}\cdot\vec{r}_{i,j})\nonumber\\
    a_{i1,j2}=\frac{1}{N}\sum_{\vec{k}}[\xi(k)-\eta(k)]\exp(i\vec{k}\cdot\vec{r}_{i,j}),
\end{eqnarray}
in which
\begin{eqnarray}
    \xi(k)=\frac{c_{1}\gamma(k)+c_{2}}{1+\sqrt{1-(c_{1}\gamma(k)+c_{2})^{2}}}\nonumber\\
    \eta(k)=\frac{c_{1}\gamma(k)-c_{2}}{1+\sqrt{1-(c_{1}\gamma(k)-c_{2})^{2}}}\nonumber,
\end{eqnarray}
here $c_{1}=4J_{\parallel}\Delta_{\parallel}/\sqrt{2}\lambda$,
$c_{2}=J_{\perp}\Delta_{\perp}/\sqrt{2}\lambda$,
$\gamma(k)=(\cos(k_{x})+\cos(k_{y}))/2$.

The Bosonic RVB wave function adopted in this study is given by
Gutzwiller projection of the mean field ground state into the
physical subspace satisfying the local constraint,
\begin{eqnarray}
    |G\rangle =
    P_{G}\Bigl[\sum_{i,j;\mu,\nu}&a_{i\mu,j\nu}&(b_{i\mu,\uparrow}^{\dagger}b_{j\nu,\downarrow}^{\dagger}-b_{i\mu,\downarrow}^{\dagger}b_{j\nu,\uparrow}^{\dagger})
     \Bigr]^{N/2}\Bigl|0\Bigr\rangle\nonumber\\.
\end{eqnarray}
Here $P_{G}$ denotes the Gutzwiller projection and $N$ is the number
of lattice sites. The mean field ground state contains two
dimensionless parameters, namely $c_{1}$ and $c_{2}$. In the mean
field theory, both of them are determined by the mean field
self-consistent equations. Here we regard them as two independent
variational parameters. This is the key difference between our
theory and that of \cite{Yoshioka}.

The proposed wave function Eq.(10) can describe both the magnetic
ordered and the quantum disordered state. As can be seen from
Eq.(9), as $c_{1}+c_{2}\rightarrow 1$, both $a_{i1,j1}$ and
$a_{i1,j2}$ becomes long ranged and the wave function describes a
state with antiferromagnetic long range order. On the other hand,
when $c_{1}+c_{2}$ deviates from 1, the RVB amplitudes $a_{i1,j1}$
and $a_{i1,j2}$ become short ranged and the corresponding wave
function describes a quantum disordered state. In fact,
$c_{1}+c_{2}=1$ is nothing but the Bose condensation condition in
the mean field theory.

On general grounds, we expect the interlayer pairing $c_{2}$ to
increase with $\alpha$ and the intralayer pairing $c_{1}$ to
decrease with $\alpha$. The transition between the antiferromagnetic
ordered state and the quantum disordered state is signaled by the
deviation of $c_{1}+c_{2}$ from 1. These expectations are confirmed
in the numerical calculation.

\section{III. The numerical techniques}
The Bosonic RVB wave function Eq.(10) can be studied by the standard
loop gas Monte Carlo algorithm\cite{loop}. In this algorithm, the
calculation of expectation value of a physical quantity
$\hat{A}$(for example the energy) is done as follows
\begin{eqnarray}
    \frac{\langle G| \hat{A}|G\rangle}{\langle G|G\rangle}
    =\frac{\sum_{\gamma,\gamma'}\psi^{*}_{\gamma}\psi_{\gamma'}\langle\gamma|\gamma'\rangle
    \frac{\langle\gamma|\hat{A}|\gamma'\rangle}{\langle\gamma|\gamma'\rangle}
    }{\sum_{\gamma,\gamma'}\psi^{*}_{\gamma}\psi_{\gamma'}\langle\gamma|\gamma'\rangle}.
\end{eqnarray}
Here $|\gamma\rangle$ denotes the valence bond basis vector and is
given by $|\gamma\rangle=\prod_{(i,j)\in\gamma}S(i,j)$.
$\psi_{\gamma}$ is the corresponding amplitude and is given by
$\psi_{\gamma}=\prod_{(i,j)\in\gamma}a_{i,j}$. The overlap between
two valence bond basis vectors $|\gamma\rangle$ and
$|\gamma'\rangle$ can be graphically interpreted as a loop gas on
the lattice by fusing the valence bonds in the two basis vectors. It
is easy to show that $\langle\gamma|\gamma'\rangle=2^{N_{L}}$, where
$N_{L}$ is the number of loops in the transition graph between
$|\gamma\rangle$ and $|\gamma'\rangle$.

As the system is bipartite, the RVB amplitude $a_{i1,j1}$ and
$a_{i1,j2}$ are in fact positive definite and the wave function
Eq.(10) satisfy the Marshall sign rule\cite{loop}. For this reason,
we can interpret
$W(\gamma,\gamma')=\frac{\psi^{*}_{\gamma}\psi_{\gamma'}\langle\gamma|\gamma'\rangle}{\sum_{\gamma,\gamma'}\psi^{*}_{\gamma}\psi_{\gamma'}\langle\gamma|\gamma'\rangle}$
as a normalized probability in the space of loop gas and can draw
samples on it with the standard Monte Carlo method. The calculation
of
$\frac{\langle\gamma|\hat{A}|\gamma'\rangle}{\langle\gamma|\gamma'\rangle}$
is easy for $\hat{A}=\vec{\mathrm{S}}_{i}\cdot\vec{\mathrm{S}}_{j}$
and the result reads
\begin{eqnarray}
    \langle \vec{\mathrm{S}}_{i}\cdot\vec{\mathrm{S}}_{j} \rangle=
    \left\{
      \begin{array}{ll}
        -\frac{3}{4}, & \hbox{$i,j \in$ same loop, different sublattices;} \\
        \frac{3}{4}, & \hbox{$i,j \in$ same loop, same sublattice;} \\
        0, & \hbox{otherwise.}
      \end{array}
    \right.
\end{eqnarray}
Thus both the energy and spin structure factor can be easily
calculated with the standard Monte Carlo procedure in the loop gas
space.

To determine the optimal value of the variational parameter $c_{1}$
and $c_{2}$, we calculate the expectation value of the energy and of
its gradients in the parameter space $(c_{1},c_{2})$ on a finite
lattice. It is useful to note that the gradients of energy can be
directly simulated by the loop gas Monte Carlo method also. Its
expression is given by
\begin{eqnarray}
    \frac{\partial E(c_{1},c_{2})}{\partial
c_{1,2}}&=&\left\langle
\Bigl(\sum_{(i,j)\in\gamma,\gamma'}\frac{\partial \ln
a_{i,j}}{\partial
c_{1,2}}\Bigr)\frac{\langle\gamma|\hat{H}|\gamma'\rangle}{\langle\gamma|\gamma'\rangle}
\right\rangle_{L}\nonumber\\
 &-&E(c_{1},c_{2})\left\langle \sum_{(i,j)\in\gamma,\gamma'}\frac{\partial \ln
a_{i,j}}{\partial c_{1,2}}\right\rangle_{L},
\end{eqnarray}
where $\langle \rangle_{L}$ denotes average over the loop gas
configurations with the weight $W(\gamma,\gamma')$.

We have used $10^{8}$ samples to calculate the energy and its
gradients to determine the optimized values for $c_{1}$ and $c_{2}$.
The boundary condition of the finite lattice is set to be periodic
in both directions. The calculation is done on a lattice with size
up to $20\times20\times2$, at which we find the critical coupling
converges to $\alpha_{c}\approx 2.62$.

\section{IV. The numerical results}
The optimized value for the parameter $c_{1}$ and $c_{2}$ as
functions of the coupling constant $\alpha$ are shown in Fig.1. As
$\alpha$ increases, the interlayer RVB pairing strength $c_{2}$
increases at the expense of the intralayer RVB pairing strength
$c_{1}$. The result is obtained on a $20\times20\times2$ lattice. It
is found that the optimized values deviate significantly from the
mean field predictions, especially for large value of $\alpha$. For
example, the mean field theory predicts that $\Delta_{\perp}$ would
reach twice the value of $\Delta_{\parallel}$ around $\alpha=4$.
However, the variational theory predicts that $\Delta_{\perp}$ is
slightly smaller than $\Delta_{\parallel}$ around $\alpha=4$. Thus,
the mean field theory overestimates greatly the tendency to form
interlayer dimer at large $\alpha$.

\begin{figure}[h!]
\includegraphics[width=8cm,angle=0]{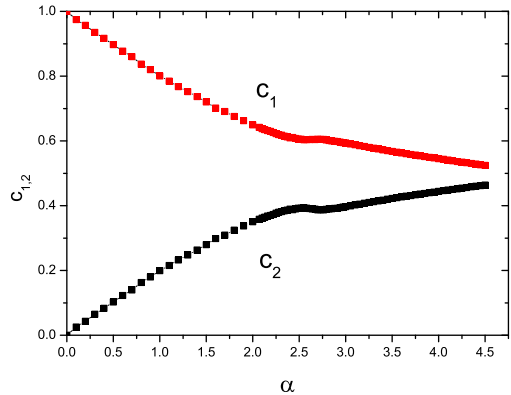}
\caption{The optimized value for the intralayer and interlayer RVB
pairing strength $c_{1}$ and $c_{2}$ as functions of the coupling
constant $\alpha$.} \label{fig1}
\end{figure}

To better understand the evolution of the variational parameters as
functions of $\alpha$, we plot the value of the $a=c_{1}+c_{2}$ as a
function of $\alpha$ in Fig.2. As we have shown above, the quantum
phase transition between the magnetic ordered state and the quantum
disordered state in our variational theory is solely controlled by
the value of $a$. The value of $a$ is seen to deviate from unity
around $2.6$, at which the Bose condensate of the spinon is gone.

\begin{figure}[h!]
\includegraphics[width=8cm,angle=0]{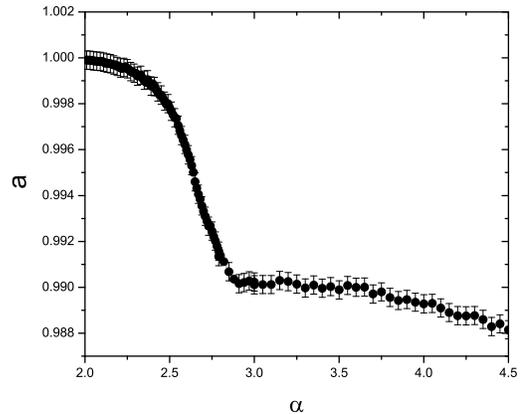}
\caption{The optimized value for the variational parameter
$a=c_{1}+c_{2}$, which controls the order-disorder transition of
BHM. } \label{fig2}
\end{figure}

To further characterize the quantum phase transition and determine
the value of the critical coupling $\alpha_{c}$, we study the
following three kinds of quantities: the spin structure factor at
the ordering wave vector, the fidelity susceptibility of the ground
state and the Binder moment ratio $Q_{2}$.

\subsection{A. Spin Structure Factor}
For a finite system, the spontaneous magnetization can be defined in
a spin rotational invariant way as the square root of the spin
structure factor at the magnetic Bragg vector. For BHM, the Bragg
vector is $\vec{Q}=(\pi,\pi,\pi)$. The spin structure factor is
defined as
\begin{eqnarray}
    S(\vec{q})=\frac{1}{N}\sum_{i,j}\langle\vec{\mathrm{S}}_{i}\cdot\vec{\mathrm{S}}_{j}\rangle\exp(i\vec{q}\cdot \vec{r}_{i,j})
\end{eqnarray}
For $q=Q$, we have
\begin{eqnarray}
    M^{2}=NS(\vec{Q})=\sum_{i,j}(-1)^{i-j}\langle\vec{\mathrm{S}}_{i}\cdot\vec{\mathrm{S}}_{j}\rangle
\end{eqnarray}
In the quantum disordered state, as the spin correlation length is
finite, $S(\vec{Q})$ is of order one. However, in the magnetic
ordered state, $S(\vec{Q})$ should scale like $N$ and thus $M$ is an
extensive quantity.

The result of the spin structure factor for a $20\times20\times2$
system is shown in Fig.3. An order-disorder transition can be seen
around $2.5$. However, the signature of phase transition in the spin
structure factor is not sharp enough for an accurate determination
of the critical coupling strength. The transition is rounded into a
crossover as a result of the finite size effect. For this reason, we
need some other quantities that are more sensitive to the transition
to determine the critical coupling.

\begin{figure}[h!]
\includegraphics[width=8cm,angle=0]{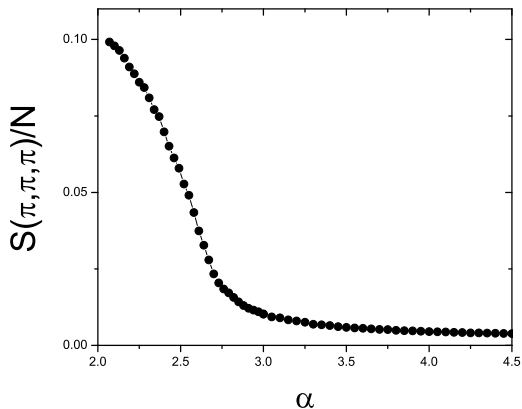}
\caption{The spin structure factor at the antiferromagnetic ordering
wave vector as a function of $\alpha$ for a system with $L=20$.}
\label{fig3}
\end{figure}

\subsection{B. Fidelity Susceptibility}
The concept of fidelity susceptibility is introduced to describe the
sensitivity of the ground state to the variation of the parameters
in Hamiltonian\cite{fidelity} and is expected to reach its maximum
at the critical coupling of a quantum phase transition, where the
ground state is the most susceptible to the variation of the
controlling parameters of the phase transition. The fidelity
susceptibility is defined in the following manner for a system with
only one parameter $\alpha$,
\begin{equation}
\chi_{f}=-2\lim_{\delta\alpha\rightarrow 0}\frac{\ln
|O(\alpha,\delta\alpha)|}{(\delta\alpha)^{2}},
\end{equation}
in which
$O(\alpha,\delta\alpha)=\langle\Psi_{\alpha}|\Psi_{\alpha+\delta\alpha}\rangle$
denotes the overlap between the normalized ground state vector for
parameter value $\alpha$ and $\alpha+\delta\alpha$.

In our variational theory, the fidelity susceptibility can be
calculated directly. We first fit the optimized variational
parameters as functions of $\alpha$ and then calculate the overlap
between variational ground states for nearby values of $\alpha$. The
overlap between the Bosonic RVB states is calculated in the
following way.
\begin{equation}
\frac{\langle\Psi|\Psi'\rangle}{\langle\Psi|\Psi\rangle}=\frac{\sum_{\gamma,\gamma'}\psi^{*}_{\gamma}\psi_{\gamma'}\langle\gamma|\gamma'\rangle
    \frac{\psi'_{\gamma'}}{\psi_{\gamma'}}
    }{\sum_{\gamma,\gamma'}\psi^{*}_{\gamma}\psi_{\gamma'}\langle\gamma|\gamma'\rangle}.
\end{equation}
In our calculation, we have set $\delta\alpha=0.01$. The result for
the fidelity susceptibility for systems of several sizes are shown
in Fig.4. A pronounced peak appears around $\alpha=2.6$. Fig.5 shows
the peak position extracted from Fig.4 as a function of the system
size $L$. It is found that the peak position converges rapidly to
its thermodynamic limit value $\alpha_{c}\approx2.62$ when $L>10$.
\begin{figure}[h!]
\includegraphics[width=8cm,angle=0]{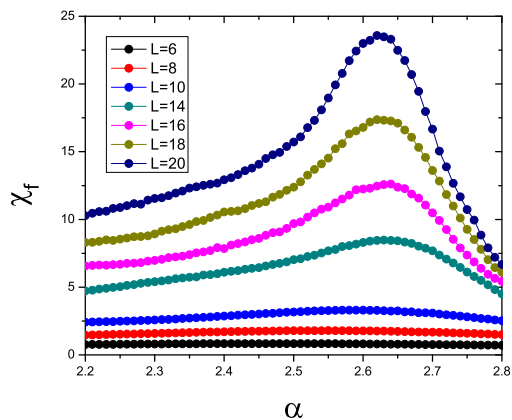}
\caption{The fidelity susceptibility for system of size
$L=6,8,10,14,16,18$ and $20$ as functions of $\alpha$. The finite
difference used to calculate the fidelity susceptibility is
$\delta\alpha=0.01$.} \label{fig4}
\end{figure}

\begin{figure}[h!]
\includegraphics[width=8cm,angle=0]{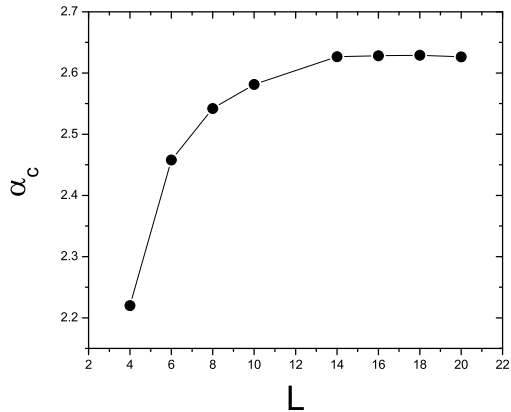}
\caption{The critical coupling strength $\alpha_{c}$ determined from
the peak position of the fidelity susceptibility. The peak position
is seen to converge rapidly to its thermodynamic limit when $L>10$.
The error bars are smaller than the size of the symbols.}
\label{fig5}
\end{figure}

\begin{figure}[h!]
\includegraphics[width=8cm,angle=0]{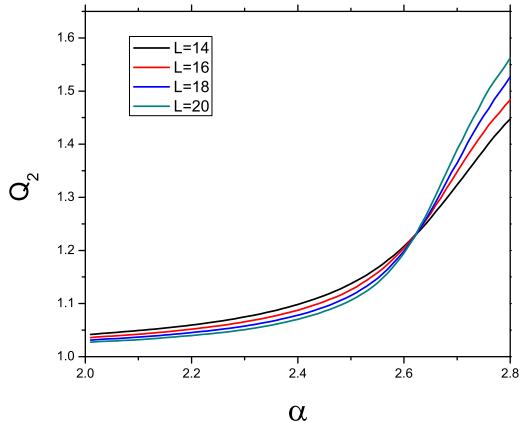}
\caption{The Binder moment ratio $Q_{2}$ for $L=14,16,18$ and $20$
as functions of $\alpha$.} \label{fig6}
\end{figure}

\subsection{C. Binder Moment Ratio $Q_{2}$}
To confirm the result derived from the fidelity susceptibility, we
calculate the Binder moment ratio $Q_{2}$\cite{Wang,Binder}. The
Binder moment ratio $Q_{2}$ is a dimensionless quantity defined in
the following manner,
\begin{equation}
Q_{2}=\frac{\langle \hat{S}_{Q}^{2} \rangle}{\langle \hat{S}_{Q}
\rangle^{2}},
\end{equation}
in which $\hat{S}_{Q}=\sum_{i,j}(-1)^{i-j}\vec{S}_{i}\cdot
\vec{S}_{j}$. Note our definition of $Q_{2}$ is slightly different
from the standard one in that it is defined in a spin rotational
invariant way, while in the standard definition only the
$z$-component of the moment is used. The Binder moment ratio is very
useful in the analysis of the critical properties as it is universal
near the critical point. More specifically, it can be expressed as a
universal scaling function of $tL^{1/\nu}$, where
$t=(\alpha-\alpha_{c})$ and $\nu$ is the critical exponent for
correlation length.

The results of $Q_{2}$ for system with $L=14,16,18$ and $20$ are
shown in Fig.6. It is found that all curves cross with each other at
approximately the same value of $\alpha$, in accordance with the
scaling hypothesis. The estimated value of the critical coupling
strength is $2.62$, in good agrement with that estimated from the
fidelity susceptibility data. The $Q_{2}$ value at the crossing
point is found to be approximately $1.23$, close but smaller than
the value(1.29) estimated from the quantum Monte Carlo simulation
with the standard definition of $Q_{2}$. Such a difference may be
caused by the difference in the definitions of $Q_{2}$.

\begin{figure}[h!]
\includegraphics[width=8cm,angle=0]{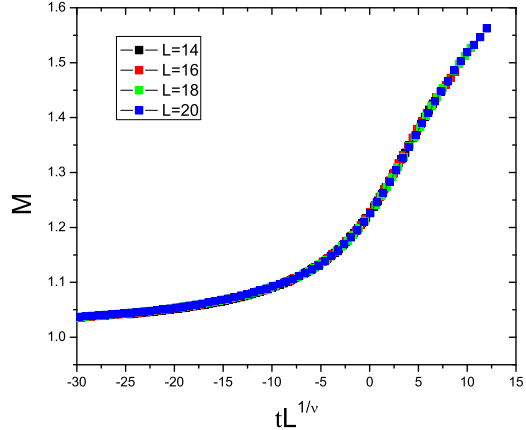}
\caption{Scaling of the Binder moment ratio $Q_{2}$ data for
$L=14,16,18$ and $20$. Here $t=\alpha-\alpha_{c}$} \label{fig7}
\end{figure}

Fig.7 shows the scaling of the $Q_{2}$ data with the scaling form
$Q_{2}=M(tL^{1/\nu})$, where $t=\alpha-\alpha_{c}$ and $\nu$ is the
exponent for the correlation length. The best fit is reached by
$\alpha_{c} \approx 2.62$ and $1/\nu \approx 1.4$. The critical
exponent so obtained $\nu\approx 0.714$ is thus quite close to the
result of the quantum Monte Carlo simulation.

\section{V. Conclusion}
In this work, we proposed a Bosonic RVB wave function with the form
of the Gutzwiller projected Schwinger Boson mean field ground state
for the BHM. We find the proposed wave function predicts a
continuous phase transition between the antiferromagnetic ordered
state and the quantum disordered state. To determine the critical
coupling strength, we have calculated the spin structure factor, the
fidelity susceptibility and the Binder moment ratio $Q_{2}$. Through
finite size scaling analysis of the latter two quantities, we find
the critical coupling to be given by $\alpha_{c}\approx2.62$, in
good agreement with the quantum Monte Carlo simulation results. The
scaling analysis of $Q_{2}$ also provides an estimate of the
correlation length critical exponent($1/\nu\approx1.4$), which is
also in good agreement with the result of quantum Monte Carlo
simulation. We find the intralayer correlation is quite large at the
phase transition point and it dominates over the interlayer
correlation for $\alpha$ twice as large the critical coupling
strength. Thus, the phase transition has nothing to do with the
dimerization instability.

Our work indicates that the Bosonic RVB wave function derived from
Gutzwiller projection of the Schwinger Boson mean field ground state
has the potential to capture the physics of quantum phase transition
with high accuracy. The failure of it in previous variational study
\cite{Yoshioka} can be attributed to the weakness of the mean field
theory, which overestimate the tendency of the system to form
interlayer dimers. Such a overestimation is closely related to the
relaxation of the local constraint in the mean field treatment,
which prohibit multiple occupation of dimer on a given bond, even if
the mean field theory points to the tendency of Bose condensation of
of such interlayer dimers. The same instability also cause the
failure of the mean field theory itself for large $\alpha$. Hence,
the form the ground state predicted by the mean field theory is
correct, however, the quantitative relation between mean field order
parameters is less meaningful. The local constraint is thus
indispensable for a correct description of the quantum
antiferromagnet with the Bosonic RVB state.

In this work, we have proved the usefulness of the variational
approach to the quantum phase transition in BHM. However, a more
detailed study of the critical behavior and the excitation spectrum
around the critical point is obviously needed to further
characterize the quantum critical point in this system. We will
leave this task to future investigations.

This work is supported by NSFC Grant No. 10774187, National Basic
Research Program of China No.2007CB925001 and and No. 2010CB923004.
The authors acknowledge the discussion with Yizhuang You on fidelity
susceptibility.

\bigskip

\end{document}